\documentclass[reprint,superscriptaddress,showframe,nofootinbib,amsmath,amssymb,aps,prd,noeprint,twocolumn]{revtex4-2}

\usepackage{graphicx}
\usepackage{dcolumn}
\usepackage{bm}
\usepackage{hyperref}
\usepackage{orcidlink}

\newcommand{\eqn}[1]{Eq.~(#1)}
\newcommand{\fg}[1]{Fig.~#1}

\begin{document}


\title{Cosmic shear with one component and its application to future radio surveys}%

\author{Yu-Hsiu Huang\orcidlink{0000-0002-4982-0208}}
\affiliation{Department of Astronomy, University of Arizona, Tucson, Arizona 85721, USA}

\author{Elisabeth Krause\orcidlink{0000-0001-8356-2014}}%
\affiliation{Department of Astronomy, University of Arizona, Tucson, Arizona 85721, USA}
\affiliation{Department of Physics, University of Arizona, Tucson, Arizona 85721, USA}

\author{Tim Eifler\orcidlink{0000-0002-1894-3301}}
\affiliation{Department of Astronomy, University of Arizona, Tucson, Arizona 85721, USA}

\author{Gary Bernstein\orcidlink{0000-0002-8613-8259}}
\affiliation{Department of Physics and Astronomy, University of Pennsylvania, Philadelphia, Pennsylvania 19104, USA}

\author{Jiachuan Xu\orcidlink{0000-0003-0871-8941}}
\affiliation{Department of Astronomy, University of Arizona, Tucson, Arizona 85721, USA}

\author{Eric Huff\orcidlink{0000-0002-9378-3424}}
\affiliation{Jet Propulsion Laboratory, California Institute of Technology, Pasadena, California 91109, USA}

\author{Pranjal R. S.\orcidlink{0000-0003-3714-2574}}
\affiliation{Department of Astronomy, University of Arizona, Tucson, Arizona 85721, USA}

\date{\today}

\begin{abstract}
We present a new approach to measuring cosmic shear: the one-component kinematic lensing (KL) method. This technique provides a simplified implementation of KL that reduces shape noise in weak lensing (WL) by combining kinematic information with imaging data, while requiring less observational effort than the full two-component KL. We perform simulated likelihood analyses to assess the performance of the one-component KL and demonstrate its applicability to future radio surveys. Our forecasts indicate that, for radio surveys, the one-component KL is not yet competitive with traditional WL due to the shallow redshift distribution of \textsc{Hi}-selected galaxies. However, when applying this method to deeper spectroscopic surveys with stronger emission lines, the one-component KL approach could surpass WL in constraining power, offering a promising and efficient pathway for future shear analyses.
\end{abstract}

\maketitle

\section{Introduction}
Weak lensing (WL) is the deflection of photon trajectories by the tidal field (see, e.g., \citep{Kilbinger2015, Mandelbaum2018} for reviews). It probes the integrated gravitational potential without assumptions about the mass-to-light ratio, making WL a powerful tool for studying large-scale structure unbiasedly, and hence provides a clean approach to understand the nature of dark energy and dark matter. However, several challenges limit the precision of WL, including uncertainties in photometric redshifts and galaxy shape measurements (see \citep{Mandelbaum2018} for details). A key limitation arises from the shape-shear degeneracy, which describes our inability to distinguish between a galaxy's intrinsic shape and the distortion induced by lensing. Since the intrinsic shape is unknown for individual galaxies, WL analyses rely on ensemble averages over large samples. The corresponding uncertainty of the intrinsic shape, known as shape noise, is around \(\sigma_\epsilon^{\rm WL} \approx 0.3\)  per galaxy \citep{Gary2002}, setting a fundamental limit on WL noise. 

Kinematic lensing (KL) is a recently developed WL technique that mitigates this limitation by combining imaging and spectroscopic information to break the shape–shear degeneracy \citep{Huff2013, Xu2023, RS2023, Xu2024, RS2024}. Because gravitational lensing distorts images but not photon frequencies, the measured velocity field of a disk galaxy retains information about its intrinsic shape. 
The observed line-of-sight rotation velocity in combination with the Tully-Fisher relation \citep{TF1977} and galaxy photometry allows us to infer the disk inclination of the unlensed galaxy and thus the first shear component. Spatially resolved kinematics also enable us to measure the misalignment between the photometric and kinematic axes  which is a signature of the second shear component \citep{Blain2002, Gurri2021}. Simulation-based studies have shown that KL can reduce the shape noise by an order of magnitude compared to traditional WL \citep{Xu2023, RS2023}, and an application to cluster lensing has demonstrated comparable gains \citep{RS2024}.

A full implementation of KL requires images and spatially-resolved spectra. Recent studies have been actively exploring the feasibility of a KL program with optical surveys \citep{Xu2023, Xu2024}. Unlike optical telescopes that need at least two visits to obtain images and spectra, radio telescopes naturally deliver both spectra and images in one visit. Although the existing radio surveys are limited to pencil-beam coverage, the upcoming wide-area radio surveys such as the Deep Synoptic Array (DSA-2000) \citep{DSA-white} and the Square-Kilometer Array (SKA) are competitive cosmic shear dataset \citep{Harrison2016}. These future radio surveys offer a potential platform for large-scale KL analyses without the need for separate imaging and spectroscopic observations.

Although the limited spatial resolution of radio telescopes makes it challenging to reconstruct a velocity map for individual galaxies. Nevertheless, the measured emission line widths can be used to determine the line-of-sight velocity with high precision. This motivates the one-component KL, where we infer the disk's inclination and consequently only one of the shear component. This simplified approach sacrifices one shear component, but retains the ability to break the shape-shear degeneracy without a spatially resolved velocity map. While motivated by radio observations, the one-component KL framework applies broadly to any dataset that measures galaxy rotation velocities without producing velocity maps.

In this paper, we develop the formalism of the one-component shear analysis and investigate its potential through simulated likelihood analyses assuming the SKA phase 2 (SKA2) survey configuration. This paper is organized as follows. We derive the data vector and the covariance matrix for the one-component shear two-point statistics in Sec.~\ref{sec:theory}. Sec.~\ref{sec:samples} describes the galaxy sample and the estimation of the number density for the simulated analyses. In Sec.~\ref{sec:forecast}, we present the likelihood forecasts for the SKA2. We discuss the implications of the forecast and possible extensions of the one-component KL in Sec.~\ref{sec:discussion}, and conclude in Sec.~\ref{sec:conclusion}.

\section{Theory}\label{sec:theory}
We define two coordinate systems for later use. The galaxy frame is aligned with the galaxy's intrinsic morphology, with the $x$-axis along the galaxy's major axis. The image frame corresponds to the sky coordinate, with $x-$ and $y-$ axes on the plane of the sky. The transformation of shear between the two coordinates is a rotation by twice the galaxy's position angle. We denote the shear components in the galaxy frame as $(\gamma_+, \,\gamma_\times)$, where $\gamma_+$ is the shear component measured at galaxy major axis and $\gamma_\times$ is measured at 45$^\circ$ from it. In the image frame, the two components are $(\gamma_1,\,\gamma_2)$, where $\gamma_1$ is aligned with the sky $x$-axis and $\gamma_2$ is at 45$^\circ$ from it. 

\begin{figure*}
    \centering
    \includegraphics[width=\textwidth]{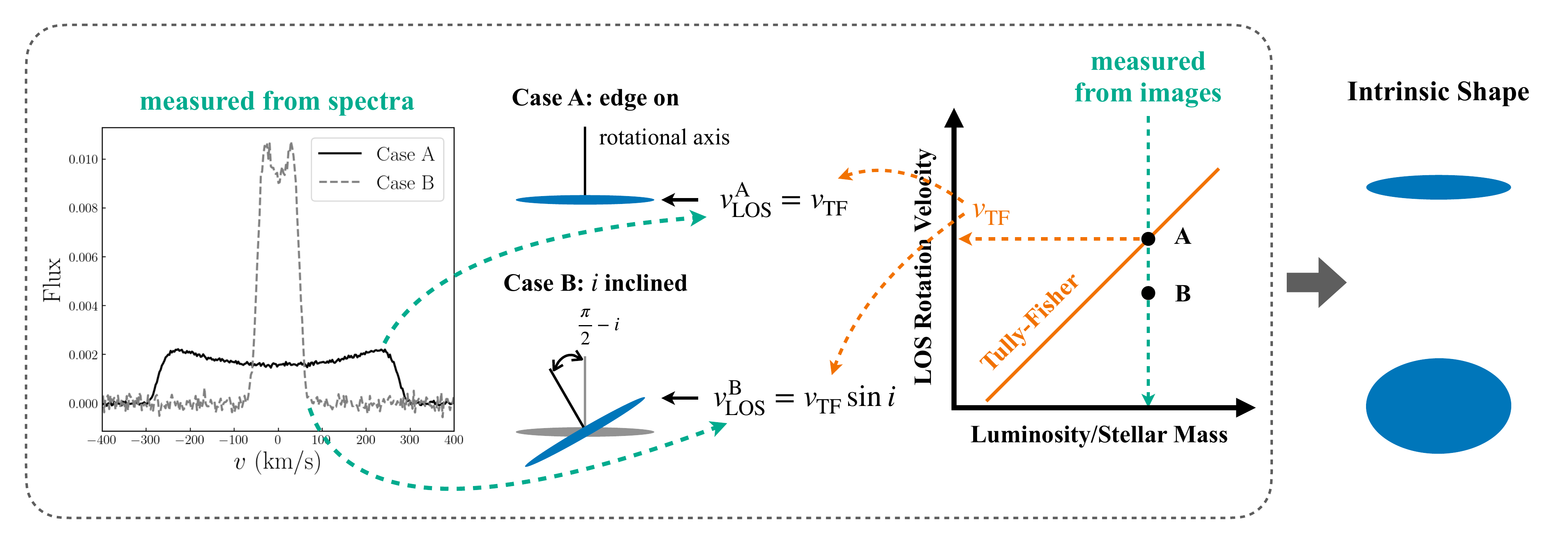}
    \caption{An illustration of the one-component KL concept. For each disk galaxy, we infer the rotation velocity $v_{\rm TF}$ from its photometry and the Tully-Fisher relation. Meanwhile, we measure the line-of-sight velocity $v_{\rm LOS}$ from the line width of the low-resolution spectrum. The ratio between $v_{\rm TF}$ and $v_{\rm LOS}$ is the disk inclination. By assuming a thin-disk model, we use this estimated inclination angle to infer the intrinsic shape of the galaxy; hence, reducing the shape noise per galaxy.} 
    \label{fig:illustration}
\end{figure*}

\subsection{One-component shear} \label{sec:one-comp}
From a spectrum with limited spatial resolution, we can infer the line-of-sight rotation velocity from the broadening of the line profile (middle panel of \fg{\ref{fig:illustration}}), assuming the gas motion is dominated by disk rotation. Combining the galaxy photometry measured from imaging data with the Tully-Fisher relation allows us to estimate the galaxy's intrinsic circular velocity (left panel of \fg{\ref{fig:illustration}}). The difference between the measured and the predicted velocities yields the disk inclination angle, which provides a constraint on $\gamma_+$ (see Eqs.~(16) and (17) in \citep{Huff2013}). However, the inclination alone can not put any constraint on $\gamma_\times$, which requires line-of-sight velocity measured at the galaxy's minor axis (see Eq.~(20) in \citep{Huff2013}). In the following, we derive expressions for the two-point correlation functions when only one shear component is measured.

Given a disk galaxy with position angle $\phi$, the cosmic shear at its location is $\boldsymbol{\gamma}=(\gamma_1, \gamma_2)^{T}$ in the image frame. We first transform the shear to the galaxy frame. Since we only measure $\gamma_+$, we set $\gamma_\times=0$, and then transform it back to the image frame. The observed one-component shear at this location, $\boldsymbol{\gamma}^{\rm one}$, amounts to
\begin{align}
    \boldsymbol{\gamma}^{\rm one} &= \mathcal{R}(-2\phi)\begin{pmatrix}
    1 & 0 \\
    0 & 0 
    \end{pmatrix} \mathcal{R}(2\phi) \boldsymbol{\gamma} \nonumber\\
    & = \begin{pmatrix}
        \gamma_1\cos^2 2\phi - \gamma_2 \cos2\phi \sin2\phi \\
        -\gamma_1\cos2\phi\sin2\phi + \gamma_2\sin^2 2\phi
    \end{pmatrix}, \label{eq:one-comp}
\end{align}
where $\mathcal{R}$ is the two-dimensional rotation matrix in the image frame. Assuming disk galaxies are randomly oriented, averaging over all position angles gives an ensemble one-component shear of $\boldsymbol{\gamma}^{\rm one} = \frac{1}{2} \boldsymbol{\gamma}$.

For two galaxies A and B separated by an angular distance $\theta$, the one-component shear correlation function is
\begin{equation}\label{eq:CF}
    \xi^{\rm one}_\pm (\theta) = \langle{\gamma^{\rm one}_1}^A {\gamma^{\rm one}_1}^B \rangle (\theta) \pm \langle{\gamma^{\rm one}_2}^A {\gamma^{\rm one}_2}^B \rangle (\theta),
\end{equation}
where the ensembles average over the position angles of the two galaxies. 
Eq.~\ref{eq:CF} simplifies to 
\begin{equation} \label{eq:one-comp-CF}
    \xi^{\rm one}_\pm = \frac{1}{4}\xi_\pm.
\end{equation}

The prefactor 1/4 corresponds to equal weighting for all galaxies in the sample. However, for a given galaxy sample, we in practice can optimize the galaxy weights based on the relative position angles in the two-point correlation function estimator, thereby boosting $\gamma_+$ and hence $\xi^{\rm one}_\pm$. We outline the corresponding derivations in Appendix~\ref{app}.

\subsection{Power spectra and covariances} \label{sec:fourier}
Within the Limber approximation, the shear angular power spectrum reads
\begin{equation}
    C_{\kappa\kappa}^{ij}(\ell) = \displaystyle \int d\chi \frac{q_i(\chi)q_j(\chi)}{\chi^2} P\left(\frac{\ell}{f_K(\chi)}, z(\chi) \right), \label{eq:power-spectrum}
\end{equation}
where $q_i(\chi)$ is the lensing efficiency
\begin{equation} \label{eq:lens-efficiency}
    q_i(\chi) = \frac{3H_0^2\Omega_{\rm m}}{2c^2 \bar{n}_i} \frac{\chi}{a(\chi)} \displaystyle \int d\chi^\prime \; n_i(\chi^\prime) \frac{f_K(\chi^\prime - \chi)}{f_K(\chi^\prime)},
\end{equation}
$n_i(\chi)$ is the source galaxy number density at $\chi$ of the $i$th tomographic bin, $\bar{n}_i$ is the mean density at the same bin, and $f_K(\chi)$ is the comoving angular diameter distance, and $P(\ell/f_K(\chi),z(\chi))$ is the 3D matter power spectrum. Following \eqn{\ref{eq:one-comp-CF}}, the power spectrum for the one-component shear reads
\begin{equation}
    \mathcal{C}^{ij}_{\kappa\kappa} (\ell,\ell^\prime)^{\rm one}= \frac{1}{4}\mathcal{C}^{ij}_{\kappa\kappa} (\ell,\ell^\prime). \label{eq:one-comp-Cell}
\end{equation}

Given the survey area $\Omega$, the Gaussian covariance matrix between two power spectra $C^{ab}_{\kappa\kappa}(\ell)$ and $C^{cd}_{\kappa\kappa}(\ell^\prime)$ is given by
\begin{align} 
    \mathbf{Cov}^{abcd}_{\kappa\kappa}(\ell,\ell^\prime) &=\frac{4\pi\delta_{\ell\ell^\prime}}{\Omega \Delta\ell(2\ell+1)} \nonumber \\
    &\biggl[\mathbf{V}^{abcd}(\ell) 
    +\mathbf{M}^{abcd}(\ell) + \mathbf{N}^{abcd}\biggr], \label{eq:covariance}
\end{align}
with the cosmic variance
\begin{equation}
    \mathbf{V}^{abcd}(\ell) = C^{ac}_{\kappa\kappa}(\ell) C^{bd}_{\kappa\kappa}(\ell) + C^{ad}_{\kappa\kappa}(\ell) C^{bc}_{\kappa\kappa}(\ell),
\end{equation}
the shot noise
\begin{equation}
    \mathbf{N}^{abcd} = \frac{\sigma_\epsilon^4}{\bar{n}_a\bar{n}_b} \left[ \delta_{ac}\delta_{bd} + \delta_{ad}\delta_{bc} \right],
\end{equation}
and the mixed term of the two
\begin{align}
    \mathbf{M}^{abcd}(\ell) &= \frac{\sigma_\epsilon^2}{\bar{n}_a} \left[\delta_{ac}C^{ac}_{\kappa\kappa}(\ell) + \delta_{ad}C^{ad}_{\kappa\kappa}(\ell)\right] \nonumber \\
    &+ \frac{\sigma_\epsilon^2}{\bar{n}_b} \left[\delta_{bc}C^{bc}_{\kappa\kappa}(\ell) + \delta_{bd}C^{bd}_{\kappa\kappa}(\ell)\right].
\end{align}

In the one-component scenario, the shape noise depends only on the single-component KL ellipticity dispersion, and the power spectrum amplitude is reduced by a factor of four (\eqn{\ref{eq:one-comp-Cell}}). The resulting covariance for the one-component shear correlation function becomes 
\begin{align} 
    \mathbf{Cov}^{abcd}_{\kappa\kappa}(\ell,\ell^\prime)^{\rm one} &=\frac{4\pi\delta_{\ell\ell^\prime}}{\Omega \Delta\ell(2\ell+1)} \biggl[\frac{1}{16}\mathbf{V}^{abcd}(\ell) \nonumber \\
    &+\frac{1}{8} \mathbf{M}^{abcd}(\ell) + \frac{1}{4} \mathbf{N}^{abcd}\biggr]. 
\end{align}
Due to the low source number density, we can safely assume that the shape-noise term dominates and we neglect the connected non-Gaussian and the super-sample covariance in this work.

Both Eqs.~\ref{eq:one-comp-Cell} and (\ref{eq:covariance}) demonstrate that the one-component KL formalism behaves similarly to the conventional two-component case, differing only by a factor of four in amplitude for the two-point statistics. Consequently, the existing KL framework can be straightforwardly adapted to the one-component analysis. In the following section, we perform simulated analyses to forecast the performance of one-component KL for future radio surveys.

\section{Radio samples}\label{sec:samples}
To perform our forecasts, we require both the source redshift distribution and the total galaxy number density. We take two approaches to infer these quantities. The first approach directly selects galaxies from a simulated catalog. In this work, we utilize the Tiered Radio Extragalactic Continuum Simulation (T-RECS) and its \textsc{Hi} extensions \citep{Bonaldi2018:trecs, Bonaldi2023:trecs}. The second approach extrapolates from empirical galaxy scaling relations combined with the local \textsc{Hi} mass function. In Sec.~\ref{subsec:trecs}, we describe how we construct KL and WL samples from the T-RECS catalog. The alternative approach based on scaling relations is presented in Sec.~\ref{subsec:local-relations}. 

Throughout this work, we adopt SKA2 as the reference survey configuration. SKA is an ongoing international radio project that will cover frequencies from 50 MHz to 15.4 GHz\footnote{\url{https://www.skao.int}}, enabling \textsc{Hi} studies out to cosmic dawn. With a planned survey coverage of 30,000 deg$^2$ and sub-$\mu$Jy sensitivity, SKA2 is a compelling opportunity for future cosmic shear measurements. 

\subsection{T-RECS catalog} \label{subsec:trecs}
T-RECS simulates the radio continuum of galaxies over a 25 deg$^2$ lightcone spanning $z=0-8$ by painting galaxies and active galactic nuclei (AGN) onto halos from P-Millennium \citep{Baugh2018:PM}, a dark matter-only simulation based on Planck cosmology. \citet{Bonaldi2018:trecs} demonstrated that this catalog reproduces the luminosity, the redshift, and the size distribution for both star-forming galaxies and AGN. The \textsc{Hi} emission was later incorporated by correlating \textsc{Hi} mass with star-formation rate, producing an extended catalog validated against local \textsc{Hi} observations \citep{Bonaldi2023:trecs}. 

We exclude AGN from our analysis and focus on star-forming galaxies. To account for instrumental limitations, we compute the expected sensitivity and resolution using the SKAO sensitivity calculator\footnote{\url{https://www.skao.int/en/science-users/ska-tools/493/ska-sensitivity-calculators} designed for SKA1}. We adopt the AA4 array configuration\footnote{This array configuration has 133 15-m antennas and 64 13.5-m antennas with a maximum baseline of approximately 150 km}, assume 10 mm precipitable water vapor, and an elevation angle of 45$^\circ$. The calculator outputs are rescaled to reflect the anticipated phase 2 (SKA2) performance, which is expected to achieve roughly ten times the sensitivity of SKA1 according to SKAO specifications. The resulting root-mean-square noises and beam sizes per continuum band and per line channel are listed in Table~\ref{tab:SKAO}.

\begin{table}[]
    \centering
    \begin{tabular}{l|cc}
    \hline\hline
       Parameter  & Band 1  & Band 2\\
       \hline
        Central frequency $\nu_c$ (MHz) & 797.5 & 1310 \\
        Bandwidth $\Delta\nu$ (MHz) & 435 & 720 \\
        Channel bandwidth $\delta_\nu$ (kHz) & 13.5 & 13.5 \\
        Continuum beam size $b_{\rm cont}$ (arcsec) & 0.5 & 0.5 \\
        Continuum RMS noise $\sigma_{\rm cont}$ ($\rm\mu Jy/beam$) & 0.392 & 0.192 \\
        Channel beam size $b_{\rm line}$ (arcsec) & 2.046 & 1.245 \\
        Line RMS noise $\sigma_{\rm line}$ ($\rm\mu Jy/beam$) & 32.037 & 20.182 \\
    \hline\hline
    \end{tabular}
    \caption{SKA2 sensitivity parameters. This table is generated by the SKA sensitivity calculator with adjustments matching the SKA2 anticipated performance.}
    \label{tab:SKAO}
\end{table}

\begin{figure*}[htb]
    \centering
    \includegraphics[width=\textwidth]{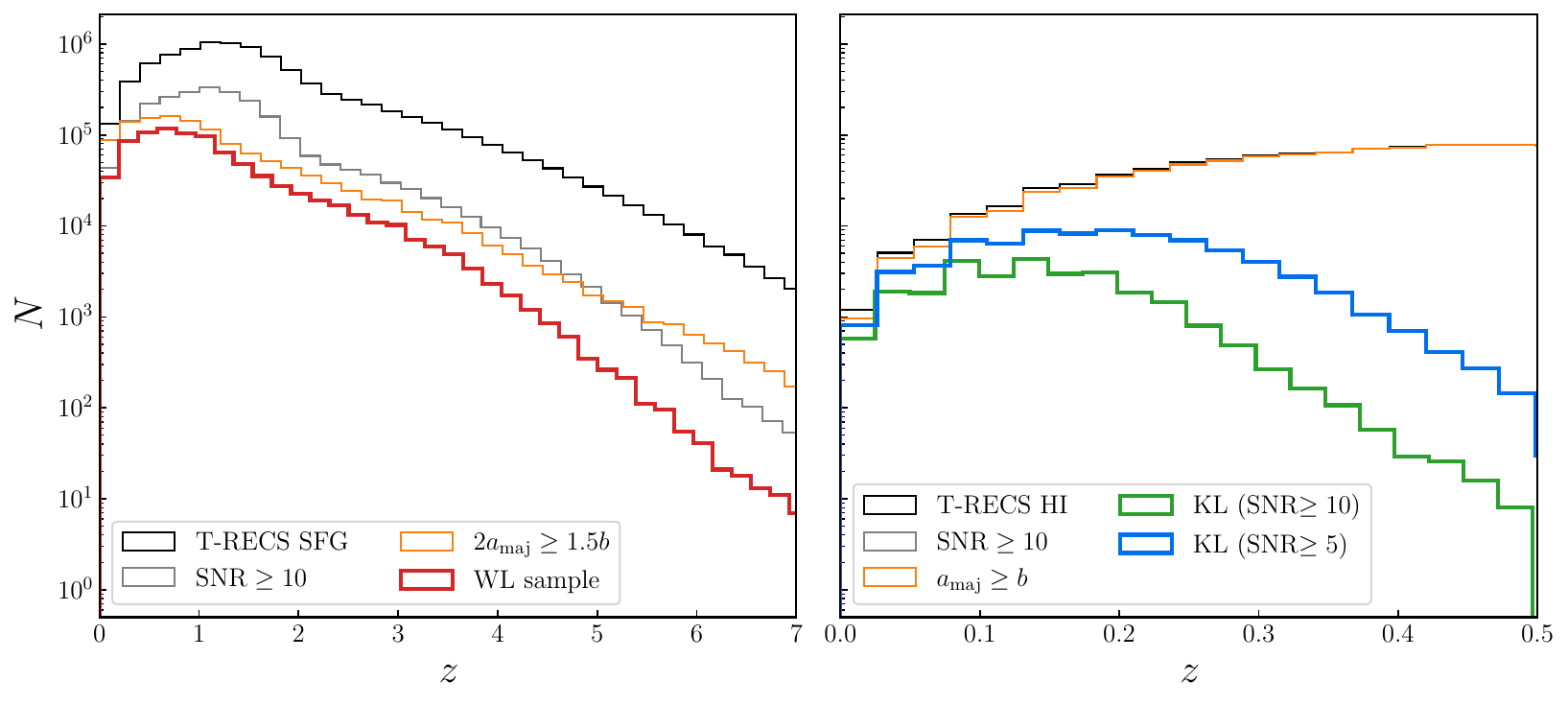}
    \caption{\textit{Left}: The redshift distribution of the WL sample (red) from the T-RECS continuum catalog. The other three histograms correspond to the distributions of the T-RECS SFGs (black), the galaxies with $\rm SNR_{cont} \geq 10$ (gray), and the galaxies with $a_{\rm maj} \geq 1.5 \,b_{\rm cont}$ (orange). \textit{Right}: The redshift distribution of the KL sample from the T-RECS \textsc{Hi} catalog. The color codes are the same as the left panel while the green histogram is the KL sample with threshold $\rm SNR=10$ and the blue histogram has threshold $\rm SNR=5$.}
    \label{fig:dndz}
\end{figure*}

\subsubsection{WL sample}
The WL galaxies should have continuum detections that are bright enough and sufficiently resolved to allow reliable shape measurements. We therefore select galaxies based on their continuum signal-to-noise ratio (SNR) and apparent size. The continuum SNR of an extended source is given as
\begin{equation}
    {\rm SNR_{cont}} = \frac{S_{\rm cont}}{\sigma_{\rm cont}\sqrt{n_b}},
\end{equation}
where $S_{\rm cont}$ is the observed continuum flux density, $\sigma_{\rm cont}$ is the rms noise per beam, and $n_b$ is the number of synthesized beams per galaxy. We approximate $n_b = a_{\rm maj}/b_{\rm cont}$, where $a_{\rm maj}$ is the apparent galaxy major-axis size of the galaxy and $b_{\rm cont}$ is the beam size in continuum observations. Both $S_{\rm cont}$ and $a_{\rm maj}$ are taken from the T-RECS catalog. Applying the selection criteria $\rm SNR_{cont} \geq 10$ and $a_{\rm maj}\geq 1.5\,b_{\rm cont}$, we obtain a WL sample spanning $z=0 -7$ with a total number density of $n_{\rm gal}^{\rm WL}=9.382\,\rm arcmin^{-2}$. The corresponding redshift distribution is shown in the left panel of \fg{\ref{fig:dndz}}.

\subsubsection{KL sample}
For the one-component KL,  we select galaxies based on the detectability of their \textsc{Hi} line emission. The SNR of a spectral line measurement is 
\begin{equation} \label{eq:SNR-line}
    {\rm SNR_{line}}=\frac{S_{21}}{\sigma_{\rm line}\delta\nu\sqrt{n_bn_c}},
\end{equation} 
where $S_{21}$ is the integrated \textsc{Hi} flux density ($\rm Jy\,km\,s^{-1}$), $\sigma_{\rm line}$ is the per-channel rms noise, $\delta\nu$ is the channel bandwidth, and $n_c$ is the number of channels across the line width. We adopt $\rm SNR_{line} \geq 10$ as our fiducial selection to ensure reliable measurement of the \textsc{Hi} line profile.

Although the one-component KL does not require spatially resolved velocity maps, we still impose $a_{\rm maj}\geq b_{\rm line}$ to avoid confusion. With $\rm SNR_{line}\geq 10$, we obtain a sample spanning $z=0-0.5$ with a total number density of $n_{\rm gal}^{\rm KL} = 0.299\,\rm arcmin^{-2}$. In Sec.~\ref{subsec:results}, we also consider a relaxed threshold of $\rm SNR_{line} \geq 5$, which increases the total galaxy number density to $0.856\,\rm arcmin^{-2}$. Both redshift distributions are shown in the right panel of \fg{\ref{fig:dndz}}.

\subsection{Local scaling relations} \label{subsec:local-relations}
As an independent validation, we estimate the redshift distribution and number density of \textsc{Hi} galaxies using empirical scaling relations. This approach serves as a consistency check for the predictions derived from the T-RECS catalog. We connect parameters in \eqn{\ref{eq:SNR-line}}, namely $S_{21}$, $W_{50}$, and $n_b$, to \textsc{Hi} mass $M_{\rm \textsc{Hi}}$. We then derive the detection threshold mass $M_{\rm \textsc{Hi}, lim}$ for a given SNR and the redshift. Integrating the \textsc{Hi} mass function up to $M_{\rm \textsc{Hi}, lim}$ yields the comoving number density of detectable galaxies, and thus the source redshift distribution.

These relations are derived from empirical scaling relations measured in the local universe. We assume no redshift evolution in either the scaling relations or the \textsc{Hi} mass function due to limited observation constraints beyond \(z\sim 0.1\). The threshold SNR is 10, and we use the local \textsc{Hi} mass function from \citet{Jones2018}. The flux-mass relation is given by \citep{Meyer2017}
\begin{equation} \label{eq:flux-mass}
    \frac{M_{\rm \textsc{Hi}}}{M_\odot} = \frac{2.36\times 10^5}{1+z} \left(\frac{d_L}{\rm Mpc}\right)^2 \frac{S_{21}}{\rm Jy\,km\,s^{-1}},
\end{equation}
where $d_L$ is the luminosity distance. The apparent size of the \textsc{Hi} disk is derived from the size-mass relation \citep{Rajohnson2022}
\begin{equation} \label{eq:size-mass}
    \log \left(\frac{D_{\rm \textsc{Hi}}}{\rm kpc}\right) = (0.501\pm 0.008) \log\left(\frac{M_{\rm \textsc{Hi}}}{M_\odot}\right) - 3.252^{+0.073}_{-0.074}.
\end{equation}
Finally, the line width is estimated from the \textsc{Hi} baryonic Tully-Fisher relation \citep{Gogate2023}
\begin{equation} \label{eq:BTFR}
    \log \left(\frac{M_{\rm bary}}{M_\odot}\right)= (3.4\pm0.2)\log \left(\frac{W^c_{50,\rm rest}}{\rm km \;s^{-1}}\right) + (1.9\pm0.6),
\end{equation}
where $W^c_{50,\rm rest} = W_{50}/(1+z)$ is the rest-frame line width and $M_{\rm bary}$ is the total baryon mass.  We convert $M_{\rm bary}$ to $M_{\rm \textsc{Hi}}$ assuming the mean \textsc{Hi}-to-baryon mass ratio of $\log \langle M_{\rm \textsc{Hi}}/M_{\rm bary}\rangle=-0.58$ \citep{Gogate2023}. 

Under SKA2 specifications, this method predicts a total galaxy number density of \(0.07\,\rm arcmin^{-2}\), about a factor of three lower than the values derived from the T-RECS catalog, with a redshift distribution peaking at $z\sim 0.1$. The overall redshift distribution is in good agreement with the T-RECS prediction. The difference in number densities likely reflects the fact that the local relations and the mass function do not account for redshift evolution, while T-RECS incorporates such evolution through modeling the observed galaxy properties. We therefore adopt the redshift distribution and number density derived from T-RECS as our fiducial samples for the subsequent forecasts.

\section{Forecast}\label{sec:forecast}
\subsection{Setup}
We perform simulated likelihood analyses to infer the parameter posteriors using \textsc{CosmoLike} \citep{Krause2017} with a Bayesian framework,
\begin{equation}
    P(\mathbf{p}\,\vert\, \mathbf{D}) \propto P(\mathbf{p}) \;\mathcal{L}(\mathbf{D}\,\vert\, \mathbf{p}), \label{eq:posterior}
\end{equation}
where $\mathbf{p}$ represents the model parameters, including the cosmological and the nuisance parameters, $\mathbf{D}$ is the data power spectrum, $P(\mathbf{p})$ is the prior, and $\mathcal{L}(\mathbf{D}\,\vert\,\mathbf{p})$ is the likelihood. The likelihood is modeled as a multivariate Gaussian,
\begin{align}
    &\mathcal{L}(\mathbf{D}\,\vert\,\mathbf{p}) \propto \nonumber\\ 
    &\exp \left(-\frac{1}{2} \left[(\mathbf{D}-\mathbf{M}(\mathbf{p}))^t \, \mathbf{C}^{-1} \, (\mathbf{D}-\mathbf{M}(\mathbf{p}))\right] \right), \label{eq:likelihood}
\end{align}
where $\mathbf{M}(\mathbf{p})$ is the model prediction evaluated at parameters $\mathbf{p}$, and $\mathbf{C}$ is the covariance matrix estimated at the fiducial cosmology. Both $\mathbf{M}(\mathbf{p})$ and $\mathbf{C}$ are computed in Fourier space as described in Sec.~\ref{sec:fourier}. We sample the posterior using the affine-invariant ensemble sampler \citep{Goodman2010} for Markov chain Monte Carlo implemented in \texttt{emcee} \citep{Foreman-Mackey2013}.

We perform forecasts for WL and the one-component KL with their respective angular power spectra and covariance matrices derived in Sec.~\ref{sec:fourier}. The galaxy redshift distribution, estimated in Sec.~\ref{subsec:trecs}, serves as the source galaxy distribution, and is divided into four tomographic bins. We use 15 logarithmically spaced multipole bins with $\ell_{\rm max}=3000$ and $\ell_{\rm min}=20$. The shape noise values are set to $\sigma_\epsilon^{\rm WL}=0.3$ for WL and $\sigma_\epsilon^{\rm KL}=0.05$ for one-component KL. 

The shape noise for one-component KL is determined by the intrinsic scatter in the Tully-Fisher relation and the precision of the line width measurement. For radio observations with a spectral resolution of approximately $5\,{\rm km/s}$, the dominant contribution to the shape noise arises from the intrinsic Tully-Fisher scatter. Based on the analysis of 189 disk galaxies, \citet{Reyes2011} estimate the scatter to be 0.05 dex, which translates to a shape noise of the order of roughly $\sigma_\epsilon^{\rm KL}\approx0.05$. 

\subsection{Results} \label{subsec:results}
In this forecast, we vary only the cosmological parameters $\mathbf{p}_{\rm c}$, keeping all nuisance parameters fixed to their fiducial values, i.e., $\mathbf{p}=\mathbf{p}_{\rm c}$ in Eqs.~\ref{eq:posterior} and (\ref{eq:likelihood}). We adopt noninformative priors for all cosmological parameters as summarized in Table~\ref{tab:priors}. We note that this systematic-free forecast favors WL since KL is immune to both intrinsic alignment and photometric redshift uncertainty.

\begin{table}[htbp]
    \centering
    \begin{tabular}{lcc}
    \hline\hline
        Cosmological parameters & Fiducial & Prior \\
    \hline
        $\Omega_{\rm m}$ & 0.3156 & [0.095, 0.585] \\
        $\sigma_8$ & 0.831 & [0.5, 1.1] \\
        $n_s$ & 0.9645 & [0.84, 1.06] \\
        $\Omega_{\rm b}$ & 0.0491685 & [0.005, 0.095] \\
        $h_0$ & 0.6727 & [0.4, 0.9] \\
    \hline\hline
    \end{tabular}
    \caption{The fiducial and prior values used in the forecast. All priors are noninformative with specifications on lower and upper boundaries.}
    \label{tab:priors}
\end{table}
\begin{figure}[]
    \centering
    \includegraphics[width=\linewidth]{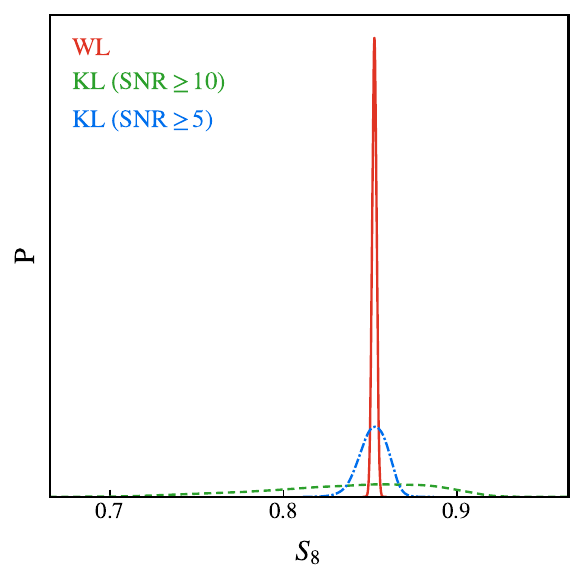}
    \caption{The marginalized posterior of $S_8$ for WL (red), KL with  $\rm SNR\geq10$ (green), and KL with $\rm SNR\geq 5$ (blue).}
    \label{fig:posterior-KL-WL}
\end{figure}

Since both WL and one-component KL probe cosmic shear, we quantify their constraining power using the width of the one-dimensional posterior on the parameter $S_8 = \sigma_8(\Omega_{\rm m}/0.3)^{0.5}$.  The marginalized posteriors are shown in \fg{\ref{fig:posterior-KL-WL}}. The corresponding 68\% confidence intervals on $S_8$ are 0.009 for WL, 0.295 for KL ($\rm SNR\geq 10$), and 0.055 for KL ($\rm SNR\geq 5$). In both cases, WL yields tighter constraints than one-component KL.

\section{Discussion}\label{sec:discussion}
\begin{figure*}[hbt]
    \centering
    \begin{minipage}{0.49\textwidth}
        \centering
        \includegraphics[width=\linewidth]{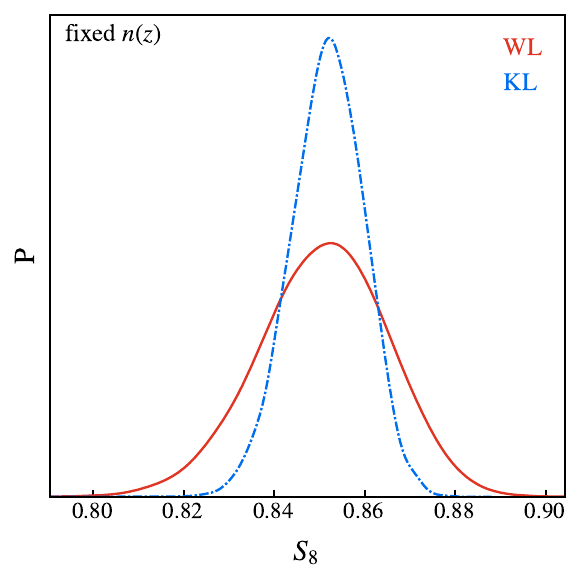}
    \end{minipage}
    \hfill
    \begin{minipage}{0.49\textwidth}
        \centering
        \includegraphics[width=\linewidth]{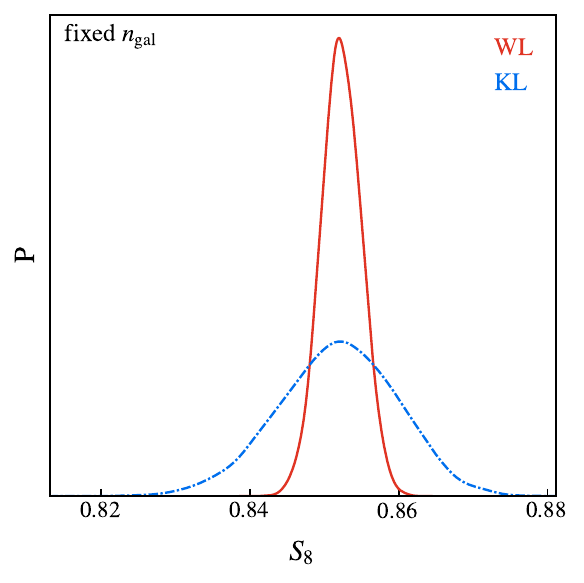}
    \end{minipage}
    \caption{\textit{Left}: Constraint on $S_8$ when fixing the source redshift distribution to the KL ($\rm SNR\geq5$) galaxy redshift distribution (the redshift histogram in the right panel of \fg{\ref{fig:dndz}}) for both WL (red) and one-component KL (blue). \textit{Right}: Constraint on $S_8$ when fixing the source number density to $0.856\,\rm arcmin^{-2}$ for both WL (red) and one-component KL (blue).  Both scenarios are done by only sampling $\Omega_{\rm m}$ and $\sigma_8$ to speed up the convergence.}
    \label{fig:dndz_ngal_comparison}
\end{figure*}

Our analysis finds that WL provides substantially tighter constraints than the one-component KL, even in the case of KL galaxies with a more relaxed threshold $\rm SNR = 5$. To understand the cause of this difference, we perform two controlled tests in which we (1) fix the source redshift distribution and (2) fix the total number density for both WL and one-component KL, respectively. In these illustrative examples, we only sample $\Omega_{\rm m}$ and $\sigma_8$.

The left panel of \fg{\ref{fig:dndz_ngal_comparison}} shows that when both methods adopt the same source redshift distribution $n(z)$, the one-component KL outperforms WL, despite the substantially lower source number density of the KL sample. This result is consistent with previous studies in the context of optical weak lensing surveys \citep{Huff2013, Xu2023}, which found that the reduction in shape noise achieved by KL provides a greater improvement than the higher galaxy number density typically available to WL analyses.

In the second comparison, we fix the total number density for both WL and KL to match that of the one-component KL $\rm SNR\geq 5$, $n_{\rm gal}=0.856\,\rm arcmin^{-2}$, while allowing each method to retain its own redshift distribution.
As shown in the right panel of \fg{\ref{fig:dndz_ngal_comparison}}, WL achieves higher constraining power than KL. 


\fg{\ref{fig:dndz_ngal_comparison}} suggests that the lens efficiency (see \eqn{\ref{eq:lens-efficiency}}) plays a more dominant role in constraining cosmological parameters than the total number density in the case of one-component KL. Since SKA2 has limited power on the detection of high-redshift \textsc{Hi} galaxies, the one-component KL source galaxies are limited to $z\sim 0.5$, whereas the WL galaxies extend to high redshifts with peak number density at $z\sim 1$. This difference in the redshift distribution leads to the result that WL outperforms the one-component KL as shown in \fg{\ref{fig:posterior-KL-WL}}. We emphasize, however, that this comparison depends on the ability of future radio surveys to reach the depths predicted by the T-RECS catalog. 

Finally, the left panel of \fg{\ref{fig:dndz_ngal_comparison}} also indicates that one-component KL has a tighter constraint on $S_8$ than WL at extremely low redshift. This suggests that one-component KL could serve as a valuable probe of the growth of structure at low redshifts where WL has poor constraining power. Comparing $\Lambda$CDM parameters constraints from WL and KL, which measure growth at different redshifts, could test deviation from $\Lambda$CDM prediction. 
Furthermore, a low-redshift KL measurement enables a joint analysis of peculiar velocity and cosmic shear data, offering a direct constraint on the growth rate that is independent of $\sigma_8$ \citep{Xu2024}.

\section{Conclusion}\label{sec:conclusion}
This work explores the potential of applying KL to the next generation of radio surveys. Owing to the limited spatial resolution of radio observations, the standard two-component KL method can not be implemented directly. Instead, we infer the disk inclination from the difference between the observed line-of-sight velocity and the circular velocity predicted by the Tully-Fisher relation, thereby measuring the $\gamma_+$ in the galaxy frame. We refer to this approach as the one-component KL method.

We develop a theoretical framework for computing a correlation function/power spectrum where only one shear component is available, and quantify the corresponding noise level. Our derivation shows that the shear angular power spectrum of one-component KL is reduced by a factor of four in amplitude compared to the two-component case, with a corresponding reduction in the Gaussian covariance matrix. Part of this loss in signal can be recovered through optimizing galaxy-pair weighting, as described in Appendix~\ref{app}.   

Using simulated likelihood analyses under a SKA2-like configuration, we compare the cosmological constraining power of WL and one-component KL based on the source population drawn from the T-RECS catalog. Our analysis considers cosmological parameters only and excludes systematic, which naturally places WL at an advantage over KL. In this systemtic-free setup, WL outperforms one-component KL primary because the WL source galaxies extend to higher redshifts. However, the inclusion of realistic systematics may alter this conclusion. For instance, intrinsic alignments and photometric redshift uncertainties would affect WL but have minimal impact on KL, as KL is insensitive to these effects.

The one-component KL methodology is also intriguing in the context of optical surveys. The main limitation of radio KL stems from the weak nature of the \textsc{Hi} emission line, while optical emission lines such as [\textsc{Oiii}], H$\alpha$, and H$\beta$ provide stronger kinematic tracers that extend to higher redshifts. These emission lines have been widely used to study galaxy kinematics for decades, making the implementation of one-component KL with optical data both straightforward and technically feasible. Using such lines could greatly enhance the constraining power of one-component KL and make it competitive to WL.

Compared to the two-component KL, the one-component method offers a substantial advantage in survey efficiency, since it requires fewer fibers or slits for spectroscopic observations. For example, a full KL analysis with DESI would require at least twice the fiber pointings per galaxy compared to a one-component KL measurement. The resulting increase in survey speed allows either wider sky coverage or deeper observations within the same survey time. 

Furthermore, removing the need for spatially resolved velocity maps enables immediate application of one-component KL to existing datasets. Current Tully-Fisher catalogs such as CosmicFlow and 2MTF already provide sufficient information to extract the $\gamma_+$ component from individual galaxies. The peculiar velocity survey of DESI will also yield ideal data for implementing the one-component KL concept. 

In summary, the one-component KL requires substantially fewer observational resources than the full two-component KL, while achieving lower shape noise per galaxy than WL. Together with the availability of suitable legacy datasets, these advantages make one-component KL a practical and promising approach for future cosmic shear analyses.

\begin{acknowledgments}
This work was supported by NASA Roman WFS 22-ROMAN22-0016. E. K., Y.-H. H. and P. R. S. were supported in part by the David and Lucile Packard Foundation. YHH thanks Dr. Ryan Keenan, Dr. Ian Harrison, and Dr. Ann Bonaldi for their valuable discussion on SKA galaxy population. YHH thanks Dr. Mike Jarvis for helpful discussions on the one-component correlation functions. This material is based upon High Performance Computing (HPC) resources supported by the University of Arizona TRIF, UITS, and Research, Innovation, and Impact (RII) and maintained by the UArizona Research Technologies department.
\end{acknowledgments}

\section*{Data Availability}
The data that support the findings of this article are openly available \citep{Bonaldi2018:trecs}.

\appendix
\section{Maximizing the one-component shear signal} \label{app}

To find the optimal weighting for the two-point estimators $\xi$, we maximize the posterior $P(\xi \,\vert \mathbf{D}) \propto P(\mathbf{D}\,\vert \xi)P(\xi)$. We assume the prior $P(\xi)$ to be uniform. Since the data is composed of multiple galaxy pair measurements, we write 
\begin{equation} \label{app:decompose}
    P(\mathbf{D}\,\vert \xi) =\prod_{i} P(\mathbf{D}_i\,\vert \xi)
\end{equation}
assuming the measurement at each pair is independent, and 
\begin{equation} \label{app:integration}
    P(\mathbf{D}_i\,\vert \xi) =\int d\boldsymbol{\gamma}^A d\boldsymbol{\gamma}^B \,P(\boldsymbol{\gamma}^A,\boldsymbol{\gamma}^B\,\vert\xi)P(\boldsymbol{\gamma}^A,\boldsymbol{\gamma}^B),
\end{equation}
where the two-point estimator averages galaxy pair $A$ and $B$. $\boldsymbol{\gamma}^{A}=(\gamma^{A}_1, \gamma^{A}_2)^T$ is the true shear of galaxy A. We further assume that the true shear is a spin-2 Gaussian field, i.e., $\boldsymbol{\gamma}^A\sim \mathcal{N}(\hat{\mathbf{\gamma}}^A, C_A)$ with the covariance 
\begin{equation}
    C_A = R(-2\phi_A)
    \begin{pmatrix}
        \sigma_+^2  & 0                 \\
        0           & \sigma_\times^2   \\
    \end{pmatrix} R(2\phi_A),
\end{equation}
where $\sigma_+$ and $\sigma_\times$ are the shape noise for each shear component and $R$ is the rotation matrix introduced in the main text.

$P(\boldsymbol{\gamma}^A,\boldsymbol{\gamma}^B\,\vert\xi)$ is a multivariate Gaussian with zero mean. If we write the data vector as $\boldsymbol{\gamma} =(\gamma^A_1, \gamma^A_2, \gamma^B_1, \gamma^B_2)^T$, the covariance will be
\begin{equation}
    C_{\rm p} = \begin{pmatrix}
        \sigma^2    & 0         & \xi_{11}  & \xi_{12}  \\
        0           & \sigma^2  & \xi_{21}   & \xi_{22}  \\
        \xi_{11}    & \xi_{21}  & \sigma^2  & 0         \\
        \xi_{12}    & \xi_{22}  & 0         & \sigma^2 
    \end{pmatrix},
\end{equation}
where $\xi_{ij}=\langle \gamma^A_i \gamma^B_j\rangle$. Note that the conventional two-point estimators have $\xi_{\pm}=\xi_{11}\pm\xi_{22}$ and $\xi_{12}=\xi_{21}=0$ because of parity symmetry. Furthermore, $2\sigma^2=\sigma_+^2+\sigma_\times^2$. Inserting both covariances back to \eqn{\ref{app:integration}} and re-organizing it, we obtain
\begin{equation}  \label{app:posterior}
    P(\mathbf{D}_i\,\vert \xi)= \lvert 2\pi C_i \rvert^{-1/2} \exp\left[-\frac{1}{2}\hat{\boldsymbol{\gamma}}^T C_i^{-1} \hat{\boldsymbol{\gamma}} \right],
\end{equation}
which is a multivariate Gaussian with covariance 
\begin{equation} \label{app:inv}
    C_i = C_{\rm p} + C_{AB} = 
    \begin{pmatrix}
        \sigma^2I+C_A   & \boldsymbol{\xi} \\
        \boldsymbol{\xi}& \sigma^2I+C_B
    \end{pmatrix},
\end{equation}
where
\begin{equation}
    \boldsymbol{\xi}=
    \begin{pmatrix}
        \xi_{11}& 0         \\
        0       & \xi_{22}  
    \end{pmatrix}.
\end{equation}

We can now choose the weighting by maximizing the likelihood
\begin{equation} \label{app:likelihood}
    \ln P(\mathbf{D}\,\vert \xi) \propto \sum_i w_i\ln P(\mathbf{D}_i\,\vert \xi).
\end{equation}
The extrema are found by setting the derivative of \eqn{\ref{app:likelihood}} with respect to $\xi_{11}$ and $\xi_{22}$, separately, to zeros. These give two equations:
\begin{equation} \label{app:x11}
    \sum_i w_i \left[ (C_i^{-1})_{13} - (C_i^{-1}\boldsymbol{\gamma})_1 (C_i^{-1}\boldsymbol{\gamma})_3\right] = 0,
\end{equation}
and
\begin{equation} \label{app:x22}
    \sum_i w_i \left[ (C_i^{-1})_{24} - (C_i^{-1}\boldsymbol{\gamma})_2 (C_i^{-1}\boldsymbol{\gamma})_4\right] = 0,
\end{equation}
where the subscripts correspond to the element index of the vector or matrix. The optimization described above is general for any cosmic shear measurement. 

For one-component KL, one can calculate the covariance $C_i$ by taking $\sigma_+ \to 0$. In this case, we have $\sigma^2=\frac{\sigma_\times^2}{2}$. $C_A$ thus reduces to
\begin{equation}
    C_A=\frac{\sigma_\times^2}{2}\begin{pmatrix}
        1-\cos4\phi_A   & \sin 4\phi_A \\
        \sin 4\phi_A    & 1+\cos4\phi_A
    \end{pmatrix}.
\end{equation}
Furthermore, if we define 
\begin{align}
    \boldsymbol{A}&=\sigma^2I+C_A \nonumber\\
    &=\frac{\sigma_\times^2}{2}\begin{pmatrix}
        2-\cos4\phi_A   & \sin 4\phi_A \\
        \sin 4\phi_A    & 2+\cos4\phi_A
    \end{pmatrix}.
\end{align}
Similarly, we have
\begin{equation}
    \boldsymbol{B}=\frac{\sigma_\times^2}{2}\begin{pmatrix}
        2-\cos4\phi_B   & \sin 4\phi_B \\
        \sin 4\phi_B    & 2+\cos4\phi_B
    \end{pmatrix}.
\end{equation}
Since both $\boldsymbol{A}$ and $\boldsymbol{B}$ are invertible, \eqn{\ref{app:inv}} can be decomposed to
\begin{align}
    C_i^{-1}&=\begin{pmatrix}
        \left(\boldsymbol{A} -\boldsymbol{\xi} \boldsymbol{B}^{-1}\boldsymbol{\xi}\right)^{-1} & 0 \\
        0 & \left(\boldsymbol{B} -\boldsymbol{\xi} \boldsymbol{A}^{-1}\boldsymbol{\xi}\right)^{-1}
    \end{pmatrix} \nonumber\\
    &\times\begin{pmatrix}
        I & -\boldsymbol{\xi}\boldsymbol{B}\\
        -\boldsymbol{\xi}\boldsymbol{A} & I
    \end{pmatrix}.
\end{align}
We can therefore find the optimal $w_i$ through \eqn{\ref{app:x11}} and (\ref{app:x22}).

\bibliography{main}

\end{document}